
\input harvmac
\def\np#1#2#3{Nucl. Phys. B{#1} (#2) #3}
\def\pl#1#2#3{Phys. Lett. {#1}B (#2) #3}

\def\physrev#1#2#3{Phys. Rev. {D#1} (#2) #3}

\def\prep#1#2#3{Phys. Rep. {#1} (#2) #3}

 \Title{hep-ph/9408384, SCIPP 94/21, UW/PT 94-07}
{\vbox{\centerline{Low
Energy Dynamical }
\centerline{Supersymmetry Breaking Simplified}}}
\bigskip
\centerline{Michael
Dine$^a$, Ann E. Nelson$^b$ and Yuri Shirman$^a$}
\smallskip\centerline{\it a) Santa Cruz Institute for Particle Physics}
\centerline{\it University of California, Santa Cruz, CA   95064}
\smallskip
\centerline{\it b) Department of Physics FM-15}
\centerline{\it University of Washington, Seattle, WA 98195}
\smallskip
\baselineskip 18pt
\noindent
We present a model in which supersymmetry is
dynamically broken at
comparatively low energies.  Previous efforts to
construct simple models of this sort have been hampered
by the presence of axions.  The present model,
which
exploits an observation of Bagger, Poppitz and
Randall to avoid this problem, is far
simpler than previous constructions.  Models
of this kind do not suffer from the
naturalness
difficulties of conventional supergravity models, and
make quite definite predictions for physics
over a range of scales from $100$'s of GeV to $1000$'s
of TeV. Thus ``Renormalizable Visible Sector Models''
are a  viable alternative to more
conventional approaches. Our approach also yields
a viable example of hidden sector dynamical supersymmetry
breaking.

\Date{8/94}

\noindent

\newsec{Introduction}
If supersymmetry is truly to provide a resolution
of the hierarchy problem, it is necessary that it
be dynamically broken.  Yet,
while various mechanisms for dynamical supersymmetry
breaking (DSB) are known, there does not yet exist
any particularly compelling particle physics
model.  Most models of supersymmetry
breaking assume breaking at a scale of order
$M_{int}=\sqrt{m_{3/2}M_p}$, with the gravitino mass
$m_{3/2}$ of order the weak scale, and
 simply put in soft supersymmetry-breaking
parameters by hand.  Moreover, in these theories,
the superpotential and Kahler potential cannot
be the most general compatible with symmetries.
In the context of string theory, a number of models
have been constructed.  However, explicit models which actually
do break supersymmetry have other difficulties,
such as a non-vanishing cosmological constant and
large flavor changing neutral currents.

An alternative possibility is that supersymmetry
is broken at a low scale, within a few orders
of magnitude of the weak scale.  In such a model,
gauge interactions can serve as the ``messengers'' of
supersymmetry breaking, giving rise to a high degree
of degeneracy among squarks and sleptons.  However,
past efforts to build such models have met
a number of obstacles.  The general strategy has been
to take some model which exhibits DSB, and to
gauge some global symmetry, identifying this with
a subgroup of the standard model gauge group.
However, this typically leads to difficulties with
asymptotic freedom.  In \ref\dn{M. Dine
and A.E. Nelson, \physrev{47}{1993}{1277}.}, this problem
was avoided by identifying the global symmetry
with a new gauge symmetry, carried both by
``supersymmetry breaking sector'' fields and by ``messenger sector''
fields which also carry
standard model quantum numbers.  A second problem
is the appearance of axions associated
with spontaneously broken R symmetries.  As explained
in \ref\nelsonseiberg{A.E. Nelson and N. Seiberg,
\np{416}{1994}{46}.},
the appearance of spontaneously
broken R symmetries is generic to models of
dynamical supersymmetry breaking.
The R axion in these models is not seen  in
 terrestial experiments, due to its large decay
constant.  However, it could be emitted by red
giants and supernovae, leading to unrealistic cooling rates.
To avoid this difficulty,
it seemed necessary to introduce additional gauge
groups, whose sole purpose was to give mass to the
axion.  The resulting models were quite unwieldy,
with extremely large groups and representations,
and suffered from several naturalness
and fine-tuning difficulties.

Recently, however, Bagger, Poppitz and Randall \ref\bagger{J.
Bagger, E. Poppitz and L. Randall, Johns Hopkins preprint
JHU-TIPAC-940005 (1994).}
have pointed out that the $R$ axion
is never a problem for astrophysics.  They noted that in
the framework of a supergravity theory, the $R$ symmetry
is necessarily explicitly broken, and the $R$ axion
obtains a mass of order
\eqn\axionmass{m_a^2 \sim {\vert F \vert^{3/2} \over M_p}.}
Here, $F$ is the Goldstino decay constant (the expectation
value of the $F$-component of some hidden sector field).
In models with radiative generation of squark and
slepton masses, this is typically of order $(100\ {\rm TeV})^2$,
so the axion mass is of order $10\ {\rm MeV}$ or larger.
This term originates from the expectation value of the
superpotential required to cancel the cosmological
constant; such contributions can also arise from other
dimension-5 R symmetry breaking
operators \nelsonseiberg.  This mass is large enough to suppress
the production of these particles in red giants and supernovae
\ref\axionbounds{G. Raffelt, \prep{198}{90}{1}.}.

With the R axion problem disposed of, one may be able to construct
simpler and more compelling models of dynamical supersymmetry
breaking.  This is the goal of the present work.
We will outline a general strategy for model
building, and apply it to some particular
examples.  The models we will describe will
suffer from none of the fine-tuning
problems of earlier work.  We will require
that some parameters be small, but will
argue that this is technically natural.
The phenomenology of the models will
be quite rich.  At scales comparable to
the weak scale, one will have the spectrum
of the minimal supersymmetric standard model,
with superpartner masses roughly proportional to gauge
couplings, and possibly
with an additional singlet and
other fields.  At higher scales,
however, there will be additional fields, including
a vector-like set of messenger quarks and leptons.
Finally, at a still higher scale, one will
find the supersymmetry breaking sector itself.

Probably the simplest model of dynamical
supersymmetry breaking is that based
on the group $SU(3) \times SU(2)$, and
we will illustrate our considerations with
this theory.  In the next section, we
will review some of the essential features
of this model.  In section 3, we will
gauge a $U(1)$ symmetry, and couple additional
fields to the model, allowing for feed-down of
supersymmetry breaking to ordinary fields.
We will compute the leading contributions
to squark and slepton masses.
We also discuss how this model could  be used in the hidden sector,
giving small but possibly adequate masses to the gauginos.

In section 4, we will take up the problem of $SU(2)\times
U(1)$ breaking.  We will show that this breaking will require the
introduction of additional fields into the model.  Several
examples involving additional singlet fields and
compatible with all existing constraints will be
worked out in detail.  All of these models will
require that some couplings be small
(of order $({\alpha_2 /
\pi})$ or $({\alpha_2 /
\pi})^2$).
Such numbers, of course, are not unfamiliar in the framework
of the standard model, and will be seen to be natural in the
sense of 't Hooft \ref\thooft{G. 't Hooft,   Lecture given at
Cargese Summer Inst., Cargese, France, Aug 26 - Sep 8, 1979.
Published in  Cargese Summer Inst. 1979.}.

In section 5, we will discuss some experimental signals of low energy
supersymmetry breaking, such as terrestial gravitino production.
Finally, in section 6 we present some final remarks and conclusions.
We will argue that, from perspectives such as naturalness,
these models are at least as successful as more conventional
intermediate scale theories.  Indeed, they solve the problems
of flavor changing neutral
currents far more easily
than such theories.   They thus represent, in our view,
a viable alternative to  conventional models,
and should be taken  seriously.

\newsec{Review of the 3-2 Model}

 The minimal model with calculable dynamical supersymmetry
breaking is  the 3-2 model, based on the gauge group
$SU(3) \times SU(2)$ \ref\threetwo{I. Affleck, M.
Dine, and N. Seiberg, \np{416}{1985}{557}.}.  It is natural, then, to
use this theory to construct a viable theory of DSB.  Here we will
recall some of the basic features of this model.  More detail
is presented in \threetwo\ and \bagger.
The chiral superfields of the model are denoted
by
\eqn\quantumnos{Q = (3,2)_{1/3}; \bar U= (\bar 3,1)_{-4/3};
\bar D = (\bar 3,1)_{2/3};
L= (1,2)_{-1}.}
Here, the numbers in parenthesis
refer to the $SU(3) \times SU(2)$ representation, while
the subscript refers to the quantum numbers under a global
$U(1)$ symmetry
of the model, which we will later wish to gauge. (This will require at
least one additional field to cancel the anomaly.)  We will refer to
this symmetry
 as ``messenger hypercharge,'' or simply as hypercharge,
but it should not be confused
with the usual hypercharge of the standard model.

The most general renormalizable superpotential consistent
with these symmetries is
\eqn\superpot{W= \lambda Q L \bar D.}
In addition to hypercharge, this model also has a non-anomalous
R symmetry.   In the limit of vanishing $\lambda$, the
theory has flat directions in which the gauge symmetry
is completely broken.  The spectrum consists of
massive vector multiplets, with mass of order
$g_i v$ (here $g_i$ denotes the
$SU(3)$ or $SU(2)$ gauge coupling,
and $v$ denotes a generic expectation value),
and three massless chiral multiplets.  These multiplets
may be represented by the gauge invariant combinations,
\eqn\multiplets{X_1 = \bar D Q L;\quad X_2= \bar U Q L; \quad
X_3 = \det Q \bar Q}
where, in the last expression,
\eqn\barq{\bar Q= \left ( \matrix {\bar U & \bar D} \right ).}
and the determinant is in flavor space.  The
actual massless states can be found by expanding these
fields about their vev's.  Note that $X_1$
and $X_3$ are neutral under the $U(1)$, while $X_2$ carries
charge $-2$.  If
the $SU(3)$ coupling is larger than the $SU(2)$ coupling,
this model reduces to supersymmetric QCD with three colors
and two flavors; in this theory, it is well-known that
instantons generate a superpotential (we follow the convention
of \bagger),
\eqn\wnp{W_{np}= {2 \Lambda_3^7 \over \det (\bar Q Q)}\ .}
For small $\lambda$, we expect that all
of the vev's are large, so that the gauge
couplings are effectively weak.  Thus one can analyze the theory simply
by minimizing the superpotential and the $D$-terms.
By simple scaling arguments, the vev's of the
scalar fields all scale as $v \sim \Lambda_3 /\lambda^{1/7}$.
One finds that supersymmetry is broken.  One can compute
the spectrum numerically \bagger, but
it is not hard to guess its main features.
There are three massless states.  Apart from the
Goldstino and the axion associated with the breaking
of R symmetry, the fermion
in the charged multiplet, $X_2$ remains massless -- this
is necessary to satisfy anomaly constraints.
The charged scalar and the other three real scalars gain
mass squared of order $\lambda^2 v^2$.  The vector multiplets still
have masses of order $g_i v$, but
are slightly split.

Before closing this section, it is worthwhile
to mention some other reasonably simple models which exhibit
dynamical supersymmetry breaking, which we will
use to illustrate some aspects of model building.
There are two general criteria for a model
to break supersymmetry dynamically. First,
the classical theory should not possess any flat
directions.  Second, the model should contain
global symmetries which are expected to be spontaneously
broken non-perturbatively.  Another model which satisfies
these two criteria is a theory with gauge group
$SU(5)$ with a single $\bar 5$ and $10$ \ref\sufive{I. Affleck,
M. Dine and N. Seiberg, \pl{137}{1984}{187}.}.
It is easy to see that this model possesses
no flat directions.   By an $SU(5)$ transformation, one
can always take the $\bar 5$ to have an entry in only
the first component.  But there is no way that one
can cancel the resulting $D$ term with the $10$.
This theory also contains a global $U(1)$ which can be gauged by
adding a single new field carrying the $U(1)$ to cancel the anomaly.

Unfortunately, in this model, there is no small parameter
which permits systematic computations.  As a result,
one can only guess what happens.  However, it is almost
certain that some of the global symmetry of the model
is spontaneously broken and that supersymmetry is
broken.

This model admits a set of generalizations\ \threetwo.
As an example, consider a model with gauge group
$SU(7)$, an antisymmetric tensor, $A_{ij}$, and
three $\bar 7$'s.  Before adding a superpotential, this
model possesses flat directions in which $SU(7)$
breaks to $SU(5)$ with a $\bar 5$ and $10$.  As
a result of supersymmetry breaking in this lower energy
theory, the flat directions are lifted; this cannot,
however, be nicely described in terms of an effective
superpotential. One can add a tree level potential
which lifts the flat directions,
\eqn\sufivew{W=A_{ij}\bar 7^i_1 \bar 7^j_2\ .}
The resulting theory is expected to have broken
supersymmetry with a good ground state.   Note that
the model has an $SU(2) \times U(1)^2$ global symmetry which may be
of use; for instance a $U(1)$ subgroup of the $SU(2)$ may be gauged
without the need for additional fields to cancel anomalies.

Still one other model of potential interest possesses gauge
group $SU(5)$, two $10$'s and two $\bar 5$'s and the
most general superpotential allowed by the symmetries.  Again
this model has no flat directions, and an $SU(2)\times U(1)^2$ global
symmetry.
 Unlike the previous case,
for small value of the superpotential coupling, the
ground state is completely weakly coupled, and everything
is calculable in principle.  However, actually minimizing
the potential is quite difficult.\foot{We thank E. Poppitz and
L. Randall for a discussion of their
efforts on this problem.}

\newsec{Feeding Down DSB}

Let us focus on the 3-2 model, and consider how supersymmetry
breaking might be fed down to ordinary fields.  No renormalizable
couplings to ordinary fields can appear in the superpotential (this
is true even if we add a singlet), so we will try to take advantage
of the hypercharge symmetry.  We don't want to identify this symmetry
with any conventional (global or local) symmetry of the standard
model.  One reason for this is that at one loop, the $D$ term
for this $U(1)$ receives a non-vanishing contribution.
We will estimate this $D$-term shortly.
But there is another reason, which is more generic.  It applies even
to models in which one does not  generate a low order $D$ term.
(examples of such models will be discussed later).
  Consider some
general model which breaks supersymmetry, and identify
a global $U(1)$ symmetry of the supersymmetry breaking sector
with a gauge symmetry carried by ordinary quarks
such as ordinary hypercharge
or  (a now gauged) $B-L$.  Squarks will then
gain mass, typically at one or two loops.  In the hidden sector,
$R$ symmetry is spontaneously broken, so gauginos can gain mass.
But gluinos, which do not carry the $U(1)$ charge,
can gain mass at best at one higher order in the
loop expansion than squarks.
For example,
if squark masses squared arise at two loops, gluino masses arise
at three loops.  As a result, gluino masses will probably be unacceptably
small\foot{The possibility that very light gluinos might still be allowed
has been much discussed in the recent literature, {\it e.g.} in
\ref\lightgluino{G.R. Farrar, Rutgers preprint RU-94-35,
HEP-PH-9407401 (1994), and
references therein.}.}.
This argument does not apply if one can gauge an $SU(3)$
symmetry of the supersymmetry breaking sector and identify it with color.
However, then one must consider
rather large gauge groups (which usually
 entails loss of asymptotic
freedom) or consider complicated structures such as that of \dn.

Since we will focus here on simple supersymmetry breaking sectors,
with only $U(1)$'s which can be gauged, we will adopt a different
strategy.  We will communicate supersymmetry breaking
to the ordinary fields through another set of ``messenger'' fields.
The messengers will include quarks and leptons ($q$ \& $\ell$)
which are vector-like
with respect to ordinary gauge interactions.  These vector-like quarks
and leptons couple to gauge singlet chiral fields whose scalar and auxiliary
components gain expectation values at the  same order
of perturbation theory, as a result
of their interactions with fields carrying messenger hypercharge.
  It is, of course, necessary
to make sure that messenger hypercharge is anomaly free.
As a result, ordinary squark, slepton, and gaugino
masses will be of the same order.
(Another possibility, which we will not explore further in this paper,
occurs in models where a D term for the messenger gauge
group is not generated at low order. Then the new vector-like quarks
and leptons may be able to also carry the messenger gauge group.)
$SU(2)\times U(1)$ breaking will involve couplings to (possibly
additional) singlet fields, in conjunction with the usual
radiative mechanism for generating negative Higgs mass via the top
quark Yukawa coupling.

Before going on to construct models, it is helpful to
study the Fayet-Iliopoulos D term generated for messenger hypercharge
in this model.  Its sign is  relevant
to model building efforts.
This $D$ term is easily estimated by considering in somewhat greater
detail the form of the spectrum.  For zero $\lambda$,
the 3-2 model possesses flat directions.  In these flat directions
there is one massless chiral field charged under the $U(1)$.
One can think of this in terms of the gauge-invariant object,
\eqn\chargedfield{X_2=\bar U QL.}
It is the fermionic component of this field which is the massless
fermion in the true vacuum at non-zero $\lambda$.
The complex scalar gains a mass squared of order $\lambda^2 v^2$.
Now it is tempting to compute the Fayet-Iliopoulos term by
noting that the leading, quadratic divergence, is cancelled
provided Tr(Y)=0, and then assuming the first, subleading
term is dominated by the light charged scalar.
The result is logarithmically divergent.
One might want to identify the cutoff with the masses of the
vector multiplets, some of which are charged under the $U(1)$.

In fact, this estimate is correct.  It follows from sum rules
for the spectrum in this model.  One can derive the sum rule
relevant to the present circumstance by the following considerations.
Work in terms of component fields (rather than
superfields) and choose
't Hooft-Feynman gauge.  This has the advantage that
the scalar fields appearing in the vector multiplets are then
complex fields.  Expand the superpotential about the minimum
in the form
\eqn\wexpansion{W = W_o + \gamma_{ij} \phi_i^- \phi_j^+ + \dots}
where $W_o$ denotes the part of the superpotential involving
the neutral fields, and we have explicitly exhibited the
part contributing to the masses of charged fields.
(There are three fields of charge $+2$ and four
of charge $-2$.  One can, however, project these fields
onto the zeroth order massive states).  The actual
scalar mass matrix has two pieces.
There is a piece of the form $m^2_{ij}\phi_i^* \phi_j$.
There is also a piece of the
form $m^2_{ij}\phi_i \phi_j$.  This piece, however, makes
a contribution to the $D$-term down by $\lambda^2 / g_i^2$
compared with that above.
So we can take the mass matrix to be:
\eqn\phistarphi{\left ( \matrix {M_V^2 + \gamma^{\dagger} \gamma & 0
\cr 0 & M_V^2 + \gamma^* \gamma^T } \right ).}
Note that the upper block, which gives the
masses of the fields with charge $+2$, is $3 \times 3$, while
the lower block, which gives the masses of fields of charge
$-2$, is $4 \times 4$.

Now let us examine the computation of the $D$ term.
Starting with the expression
\eqn\dtermexpression{\vev D= \left({ g_Y^2 \over 2 \pi}\right)^4
\sum_i y_i
\int {d^4 k \over k^2 + m_i^2},}
the leading divergence cancels.  The subleading
term is given by
\eqn\subleading
{\vev D= {g_Y^2 \over 16 \pi^2} \sum_i y_i m_i^2
\ln(\Lambda^2/m_i^2).}
The divergent part is easily seen to cancel, in view of the
structure of the mass matrix described above.  It is proportional
to \eqn\traces{Tr(M_V^2 + \gamma^{\dagger} \gamma) - Tr(M_V^2 +
\gamma^*
\gamma^
T)=0.}
 In eqn. \subleading, there is a
piece proportional to $m_{\ell}^2 \ln(\Lambda^2/m_{\ell}^2)$.
(Recall $m_{\ell}^2$ is the mass of the light charged field,
of order $\lambda^2 v^2$).  In view
of the cancellation of infinities we have just noted, $\Lambda$
in this term must be replaced by $M_V$, where $M_V$ is some
typical vector mass, of order $g^2 v^2$ (where $g$ is the $SU(3)$
or $SU(2)$ gauge coupling).
\foot{It is easy to check that there are no terms of order
$M_V^2 ln(M_V^2)$ or $M_V^2$.  This follows from
the form of the mass matrix in eqn. \phistarphi.}
Putting this together, we have with logarithmic accuracy that the
coefficient of the $D$ term is given by
\eqn\dterm{\xi^2= { g_Y^2 \over 16 \pi^2} m_\ell^2
\ln(g^2 /\lambda^2).}
Note that if we cancel the anomaly of this model by adding a field,
$E$, of charge $+2$, the sign here is such that this field acquires
a positive mass on account of the $D$ term, and the $D$ term
has a non-zero expectation value.

We can use this result to build models.  As explained earlier, we
would like to have a gauge singlet field obtain an expectation
value.  So we introduce the following fields, with their corresponding
U(1) charges in parenthesis\foot{An alternative charge assignment,
which allows the new $SU(3)\times SU(2)\times U(1)$ gauge groups to be
simply unified, is to take  the $P$ and $N$ fields to have charges $\pm
4$. With a suitable superpotential and additional singlets
this also leads to a satisfactory model.}
:
\eqn\pnsmodel{E(+2),P(+1),N(-1),S(0)\ .}
In addition, we include a set of vector-like quarks and leptons.
We take these to have {\it conventional} $SU(3) \times SU(2)
\times U(1)$ quantum numbers, e.g.,
\eqn\vectorlike{q(3,1)_{-2/3}, \bar q(\bar 3, 1)_{2/3},
\ell(1,2)_{1}, \ell(1,2)_{-1}.}
For the superpotential we take:
\eqn\pnsw{W= \lambda_1 PNS + {\lambda_2\over2} E N^2 + {\lambda_3 \over 3}
S^3
+ k_1 S \bar q q + k_2 S \bar \ell
\ell.}

What are the dynamics of this model?  On account of the D-term,
the field $N$ obtains an expectation value.
This leads to a mass for several of the fields.
In particular, the $S$ and $P$ fields pair to gain mass.
Note that the sign of the $D$ term is relevant here.  Had the
$D$-term had the opposite sign, some linear combination of the
$P$ and $E$ fields would have obtained  a vev.
There would have been a massless chiral field at lowest order.
At next order, if this field received a negative mass squared, it
would have gotten a vev, inducing  as well the vev for the $S$ field
which is needed to give the $q$ and $\ell$ fermions mass.
With the given sign of the D term, it appears to be more difficult to
give the $S$ field the necessary vev.  However, suppose that
the coupling $\lambda_1$ is very small.
In this case, corrections to the $S$, $P$ and $E$ masses
from gauge field exchanges (at two
loop order) can be important.  These corrections are easily
estimated.  Just as we have argued that the light charged scalar
makes the most important contribution to the D term, so
this field can be argued to make the most important contribution
here.  In ref. \dn, it was shown that the two loop contribution
of \fig\twoloops{Two loop diagrams contributing
to scalar masses in various models.  Dashed lines
denote scalar fields; wavy lines are gauge fields.
In (a) and (b), the scalar emits a gauge field
which couples, in turn, to fields without a supersymmetric
spectrum; in (c) the scalar couples to other scalars through
the $D$ term; in (d), it couples to its fermionic partner
and a gaugino.  The labeling in the figure refers
to the contributions to masses of ``ordinary'' squarks
and sleptons.  For the $P$. $N$ and $E$ fields it is
the hidden sector fields which run in the loop.}
to the mass of a field of charge $y$, to first order in the
supersymmetry breaking mass shifts,
can be written as $y^2 \tilde m^2$,
where
\eqn\twoloopsuone{\tilde m^2 = 8 \left({g_Y^2 \over 16 \pi^2}\right)^2
\sum_i (-1)^F y_i^2 m_i^2 \ln(\Lambda^2 / m_i^2).}
here the sum is over the fields appearing in the diagram;
$y_i$ are their U(1) charges and $m_i^2$ their masses.
Again, we can consider the contributions of the different
states.  The same sum rule we used before shows that the
leading divergent term cancels; the subleading terms
give a result equal to
\eqn\twoloopresult{\tilde m^2 =
-32 \left({g_Y^2 \over 16 \pi^2}\right)^2
m_{\ell}^2
\ln
\left({g_i^2\over\lambda^2}\right)}
where $m_{\ell}$ is the mass of the light charged field of the
$3-2$ model, and $\lambda$ is the coupling appearing in the
superpotential of that model.

With this correction, the full potential of the model is
\eqn\modifiedpotential{
V= \sum \left\vert {\partial W \over \partial \phi_i }\right\vert^2
+ {1\over2} g_Y^2 (\xi^2 + 2 \vert E \vert^2 + \vert P \vert^2
- \vert N \vert^2)^2 +\tilde m^2 (4 \vert E \vert^2 + \vert P \vert^2
+ \vert N \vert^2).}
Now if $\lambda_1$ is small enough, the negative mass
term will lead to vev's for $P$, $N$ and $E$; this in turn
will drive vev's for $S$ and $F_S$.  One might
imagine that $\lambda_1$ would have to be
quite small, of order $g_y /( 4 \pi)$,  but
in practice, it turns
out that $\lambda_1$ does not need to be especially small.
For example, taking
\eqn\param{g_y=1,\quad
\lambda_1=0.2,\quad \lambda_2=0.3, \quad \lambda_3=0.3,}
we find the potential minimized at\eqn\minim{n=-2.44\xi,\quad
p=1.40\xi,
\quad s=1.28\xi,\quad e=1.27\xi,\quad F_S=-0.188\xi^2\ .}
It will be convenient to work in a range of
parameters for which $F_S$ is relatively small,
($F_S\ll k_{(1,2)}{\vev S}^2$).
This is achieved, for example, if $\lambda_3$ is small.

Now that $S$ and $F_S$ have expectation values, the
stage is set to give masses to squarks, sleptons, and
gauginos.  Gaugino masses will arise through the
one loop diagrams of \fig\ugino{
One loop diagram contributing to gaugino
masses.}. To leading order in $F_S$, the resulting mass
is
\eqn\gauginomass{m_{\tilde g}= {g_i^2 \over 16 \pi^2}
{F_S \over S}.}

Squark and slepton masses squared arise at two loops, and  thus the
masses are
of the same order as gaugino masses.   Their evaluation is
somewhat more complicated, involving the same set of diagrams
as in \twoloops, where now the gauge fields are those of
conventional $SU(3) \times SU(2) \times U(1)$, rather
than those of $U(1)_Y$.    We can determine the mass again
by repeating the computation ref. \dn, working carefully to
second order in the supersymmetry breaking mass shifts.
In the present case, the fields appearing in the loops
are $q$ and $\bar q$, or $\ell$ and $\bar \ell$.A straightforward
computation
gives
\eqn\squarkssleptons{\tilde m^2 = \sum_a 2 C_F^{(a)}
\left({g^{(a)2} \over 16 \pi^2}\right)^2{F_S^2 \over S^2}.}
Here $a$ denotes the gauge group (so, for example,
quark doublets obtain a contribution from $SU(3)$,
$SU(2)$ and $U(1)$ gauge field exchange).

Note, in particular, that these contributions are {\it positive}
and that they depend only on the gauge quantum numbers
of the fields.  Note also that no Fayet-Iliopoulos
$D$ term is generated for ordinary hypercharge at low
orders.  This is because, before worrying about ordinary
squarks and sleptons, the model has a left-right symmetry
which exchanges $q$ and $\bar q$ and $\ell$ and $\bar \ell$.
As a result, the first potential contributions to the $D$
term appear at three-loop order, and are harmless.  This is a
significant improvement over the model of ref. \dn,
where equality of certain gauge couplings had to
hold to a good approximation to avoid such $D$ terms.

The $3-2$ model has particular appeal because of its
simplicity.  The approach which we have adopted here
to feeding down supersymmetry breaking to ordinary
fields is probably the simplest one available in this case.
No ``unnatural acts" were required here.  One coupling
constant had to be reasonably small, but this is perfectly
consistent with  't Hooft's notion of naturalness.

The most popular approach to communicating supersymmetry breaking is
to have supersymmetry broken at a relatively high scale, around
$10^{11}$ GeV, and then use supergravitational and other Planck scale
couplings to feed supersymmetry breaking to the visible sector. An
early attempt to use the 3-2 model in this way \threetwo\ was discarded
because ordinary gaugino masses could only arise through operators
such as \eqn\gmassop{\int d^2\theta {\CO(1)\over m_P^3}
QLD\ \tilde W_\alpha
\tilde W^\alpha \ ,} leading to gaugino masses suppressed by two
powers of $m_P$ relative to the weak scale. With  gauged hypercharge
and the additional $P,N,S$ and $E$ fields however one could have the
operator
\eqn\newgmassop{\int d^2\theta {\CO(1)\over m_P} S\tilde W_\alpha
\tilde W^\alpha\ ,} leading to  gaugino mass terms of order $\alpha_Y/\pi$
times the squark and slepton masses. This
could
be acceptable provided $\alpha_Y$ is not too small.\foot{It is
also necessary to understand the smallness of the $\mu$ parameter
in this framework.  $S H_u H_d$ must be forbidden in the
superpotential, perhaps by
a discrete symmetry acting on the Higgs.  One probably wants
this to be an R symmetry so that the coupling $S^{\dagger}H_U
H_D$ will be allowed in the Kahler potential,
generating a $\mu$ parameter
of order $m_{3/2}$.}

Alternatively, the 3-2 model could serve as a hidden sector with
communication of supersymmetry breaking  done
by ultraheavy  Grand Unified Theory (GUT)
mass fields  which couple to the field $S$ and
carry ordinary gauge quantum numbers. Such
a model would be similar to the MSSM,
however the squark, slepton and gaugino
masses would arise from ordinary
gauge interactions and be calculable
as in the visible sector model.
Degeneracy of the
squarks and sleptons could be upset by Planck scale physics, leading
to excessive flavor changing neutral
currents (FCNC) unless the ultraheavy masses are much less than
$\CO((\alpha_Y/\pi)
(\alpha_{\rm GUT}/\pi) M_P)$.

It is interesting to ask whether other supersymmetry
breaking models might give different possibilities for
communication of
supersymmetry breakdown.
We have explained
why it is probably necessary to insulate the supersymmetry
breaking sector from the visible sector (loss of asymptotic
freedom and suitable
gluino mass).  But, as we have already mentioned, there
do exist models in which it is not necessary to introduce
spectators to cancel anomalies.  Consider the $SU(7)$ model.
Here we can gauge a U(1) which rotates the $7$ and $\bar 7$
appearing in the superpotential  of eqn. \sufivew\
by opposite phases.  In this model, rather than introducing
three fields, $E$, $P$ and $N$, we can simply introduce
two fields of opposite charge, $\phi^+$ and $\phi^-$,
and a singlet, $S$, with couplings
\eqn\alternativew{\lambda_1 S \phi^+ \phi^- + \lambda_2 S^3 +
 + k_1 S \bar q q + k_2 S \bar \ell \ell.}
Here $S, q, \bar q$, etc. will play a similar role in feeding down
supersymmetry breaking as the corresponding fields in the
$3-2$ case.
Now, however, the model has a discrete symmetry which
interchanges $\phi^+$ and $\phi^-$, as well as
$\bar 7_1$ and $\bar 7_2$.  If this symmetry
is not spontaneously broken, then, because the $D$ term
for the $U(1)$ is odd under this symmetry,
there can be no Fayet-Iliopoulos term.  Because of the
strongly coupled nature of the theory, we cannot compute
the masses generated at two loops for $\phi^+$ and $\phi^-$.
However, as long as they are non-zero, we can obtain an
acceptable model.  If the masses are negative, the minimum
of the potential, for a range of parameters, has a non-vanishing
$\vev S$ and $\vev{F_S}$.  If the masses are positive, one loop corrections
induce a negative mass for $S$, which in turn leads to
non-zero $\vev{S}$ and $\vev{F_S}$.  The rest of the story of feed-down
and $SU(2) \times U(1)$ breaking then proceeds as in the
$3-2$ case.  This is, of course, not the only alternative
model, but we suspect that the strategies we have used
here are rather general.  We favor the $3-2$ model
because of its simplicity.

\newsec{Ordinary matter and $SU(2) \times U(1)$ Breaking}
The minimal ``ordinary'' sector consists of the usual fields of the minimal
supersymmetric standard model (MSSM). The famous ``mu problem'' of the
MSSM, {\it i.e.} how
to give the Higgs fields a weak scale supersymmetric mass term,
shows up here as well. We do not  include an $H_u H_d$
term in the superpotential, since our philosophy is that all masses
should arise through dimensional transmutation.   In \dn,
this problem was dealt with by introducing a singlet, $S^{\prime}$,
with couplings to the Higgs doublets, and to an additional pair
of vector-like quarks and leptons.  These extra
fields were required in order
to generate sufficiently large negative mass for the singlet.
The model had several virtues:
the superpotential could be taken to be the most general
compatible with certain discrete symmetries; there were
no new sources of CP violation beyond the KM phase,
and the model predicted an interesting set of states beyond
the MSSM at weak-scale energies.

In the present case, we might hope to break $SU(2) \times U(1)$
in a simpler fashion, exploiting the singlet $S$ we have
already introduced.  With a coupling
$\lambda_h H_dH_u S$, the vev of $S$ could
provide a ``mu-term'' type
mass for the Higgs.  However, unless there are delicate cancellations
between different terms, one cannot obtain an acceptable spectrum
this way.  In order that higgsino masses be comparable to
the $Z$ mass, we require
\eqn\lambdah{\lambda_h \vev{S}  \sim m_Z.}
In addition, there are terms in the Higgs potential
of the form \eqn\negmass{\lambda_h\vev{F_S} H_u H_d.}
These also shouldn't be too much larger than $m_Z^2$.
So we require that \eqn\fs{\vev{F_S}  /S^2 \sim \lambda_h.}
On the other hand, the two loop contribution to the
Higgs mass, eqn. \squarkssleptons, should not be much
smaller than $m_Z^2$, and this is incompatible with these
two conditions.  At best, an acceptable spectrum can be
obtained only by appreciable fine tuning.

%

While this simplest possibility seems not to
work, we have found several viable approaches to
$SU(2)\times U(1)$ breaking.
One is to include another singlet field $S'$, and
to add to the superpotential
\eqn\superspace{\lambda_h H_u H_d S^{\prime}
+ {\lambda_{S'} \over 3} S^{\prime^3}+ k_3 S^{\prime 2} S.}
Think of $k_3$ as the small parameter in this lagrangian,
while the other couplings are of order one.
For small $k_3$ and real $F_S$, the imaginary part
of $S^{\prime}$ obtains a negative mass-squared,
$k_3 F_s$.  So $S^{\prime}$ obtains a vev:
\eqn\sprime{S^{\prime 2} \sim {k_3 F_s \over \lambda^2}.}
Note that in this estimate, we can neglect the term cubic
in $S^{\prime}$; it is down by $\sqrt{k_3}$.  Note also
that $F_{S^{\prime}} \sim {k_3 \over \lambda} F_s$.
In particular, in terms of powers of $k_3$, $F_{S^{\prime}} $
is comparable
to $S^{\prime 2}$.
Now if $k_3 \sim ({\alpha_w \over \pi})^2$,
the $H_u H_d$ term in the Higgs potential is
comparable to the terms coming from top quark exchange
and
the higgsino mass is comparable to the Higgs mass.

The main problem with this idea is that $k_3$, more
or less by accident, must be of the correct order
of magnitude.  Note, however, that it is natural for
$k_3$ and the other couplings we have omitted to
be small.  For example, suppose we have an approximate
symmetry under which \eqn\discretesymm{S^{\prime} \rightarrow
e^{2 \pi i/3} S^{\prime}; \quad H_u H_d \rightarrow
e^{4 \pi i/3} H_u H_d; \quad S \rightarrow S.}
This symmetry explains the smallness of the couplings
$S S^{\prime 2}$ and $S H_u H_d$ which violate the
symmetry by the same amount, and $S^{\prime} S^2$
which violates the symmetry by a different amount.

Another fairly simple alternative, which does not require
any small nonzero couplings,  is to have $k_3$ be zero but include couplings
  \eqn\moresupersp{k_4 S'\bar q q+k_5S'\bar\ell \ell\ .}
Now a suitably small vev for $S'$
will be induced radiatively at one loop provided $S$ gets a vev
and the couplings $k_4$ and $k_5$ are sufficiently small.
This vev is easily computed; ignoring $k_5$, one finds
\eqn\tadpole{S^{\prime~3} = {1 \over 32 \pi^2} {k_4 F_S^2 \over
k_1S}.}
However it is difficult to use symmetry arguments to explain the
omitted
couplings such as $S^{\prime} S^2$, which would lead to a large
$S^{\prime}$ vev.  At best, approximate symmetries
such as those we have described earlier
tend to keep these couplings naturally
as small as $k_4$ and $k_5$, and this is perhaps barely
small enough.

A third approach, which does not require small parameters,
repeats the construction of \dn.  In addition to $S^{\prime}$,
one includes a second set of vector-like quarks and
leptons (beyond $q$, $\bar q$, $\ell$ and $\bar \ell$),
$q^{\prime}$, $\bar q^{\prime}$, $\ell^{\prime}$ and $\bar
\ell^{\prime}$.  One now introduces couplings
\eqn\sprimew{
S^{\prime}H_u H_d +S^{\prime} \bar q^{\prime} q^{\prime}+
S^{\prime} \bar \ell^{\prime}\ell^{\prime} + S^{\prime~3}
 + H_d Q \bar q^{\prime}.}
(In this expression $Q$ now denotes the conventional quark doublets.)
This structure can be enforced by a $Z_3$ symmetry
which rotates ordinary fields and primed fields
by $e^{2 \pi i/3}$.  There is a danger, here,
of strangeness changing neutral currents if
$\bar q^{\prime}$ is too strongly mixed with the
light down quarks.  This can also
be avoided by suitable (approximate) discrete symmetries.

The primed quark and lepton fields obtain two loop
masses just like the ordinary fields.  These, in turn,
lead to negative masses at one loop for the field $S^{\prime}$
and for $H_d$.  Note that, as in \dn,
these are comparable to the two loop masses
for the weak doublets, because the
fields in the loop carry color, and because
of color and logarithmic factors in the diagrams.
With all couplings of order one, one
can readily obtain an acceptable spectrum \ref\bagnasco{
J. Bagnasco, SCIPP preprint to appear.}.  Not only
is this model the most general consistent with symmetries,
but all of the phases in the superpotential, apart from
the KM phase, can be removed by field redefinitions,
so there are no new sources of $CP$ violation.

All of these approaches result
in a low energy theory which is similar to the MSSM, with $SU(2)\times
U(1)$ breaking driven by the usual top quark radiative correction,
and with an additional light
singlet. There are fewer potentially
free parameters, however, associated with
supersymmetry breaking.  In the limit that $F_S$ is small compared
with the masses in the messenger sector, the squark, slepton, and
gaugino masses depend only on $F_S/ S$,  gauge couplings, and the
ordinary $SU(3)\times SU(2) \times U(1)$ quantum numbers of $q$ and
$\ell$,
while  the higgsino and
Higgs masses depend also on $\lambda_h \vev{S'}$ and $\lambda_h F_{s'}$.
Thus with small $F_S$,
once the top mass and
weak scale are fixed  the superpartner and Higgs masses
depend
only on two
additional parameters. For any value of $F_S$, in all
versions of this model the squark and slepton masses naturally come
out degenerate, since the leading contributions come from gauge
couplings, and do not lead to new sources of  FCNC. In general the superpartner
masses also depend on
$\lambda_\ell$ and $\lambda_q$.

Besides the superpartner masses, other supersymmetry-breaking
couplings are also nonzero. For instance there will be trilinear
scalar couplings; these arise at two loops. Supersymmetry-breaking
dimension-4 and higher couplings
also arise radiatively. We believe all these supersymmetry breaking
couplings to be  too small to be phenomenologically
interesting.

We have already noted that in the third model, the only source
of $CP$-violation are the KM phase (apart from the
$\theta$-parameter).  In the first two models,
all CP violation may be removed from
the superpotential couplings in the messenger and supersymmetry
breaking sectors, except for the phase of $k_3$ in the first version, or
$k_4$ and $k_5$ in the second. Thus the low energy supersymmetry breaking
parameters will be CP conserving, except for one phase in the Higgs sector.
Fine-tuning of this phase to $10^{-2}-10^{-3}$ is required to avoid
inducing
electric dipole moments for the neutron and for atoms. The
usual strong CP problem also still exists.

As we have noted,  in the third version, where the
$q^{\prime}$ and $\ell^{\prime}$ have weak scale
masses, there is a danger of flavor changing neutral
currents.  This problem is not as severe in the first
two models, since,   provided that the couplings $k_i$ are of order
one, the masses of the $q$ and $\ell$
are of order $(16 \pi^2/g_2^2)$ times the weak
scale. Thus mixing is highly suppressed by the large masses,
and is
not a problem.  However
it would still be of interest to study the potential for observing
nonstandard FCNC and CP violation in the B meson system.

\newsec{Experimental signatures}

We believe that the model-building strategy we have
described is rather general.  There are a number of
predictions for supersymmetry phenomenology
which follow in this framework:
\item{1.} The masses of squarks and sleptons are governed
(apart from the top squark) by their gauge
couplings, in accord with eqn. \twoloopresult.
Flavor changing neutral currents are not a problem.
\item{2.}  The masses of the gauginos are related
in a well-defined way to those of squarks and sleptons,
as can be seen by comparing equations \gauginomass\
and \squarkssleptons.  For example,  when $F_S$ is small
the ratio
of squark to gluino masses is approximately
$2 \over \sqrt{3}$.
\item{3.}  The Higgs sector is necessarily more complicated
than that of the MSSM, if one insists that {\it all}
masses arise from dimensional transmutation.
  One expects at a minimum that
there is an additional  gauge singlet with weak scale mass.
\item{4.}  There is new physics at a variety of scales.
The fields $q,\bar q, \ell$ and $\bar \ell$,
as well as the fields $P,N$ and $E$ lie at a comparable
scale.  Finally, the supersymmetry breaking fields
of the $3-2$ sector lie at energies $1-2$ orders of magnitude
larger.  So one expects new physics up to scales of order
$10^4$ TeV or so.
\item{5.} The supersymmetry breaking scale is constrained from below
by the need to have the R-axion heavier than 10 MeV, and from above
by the cosmological requirement that the gravitino be lighter than
10 keV \ref\gravitino{see {\it
e.g.} T. Moroi,
H. Murayama,  and M. Yamaguchi, \pl{303}{1993}{289} and references
therein.}
and is predicted to be in the range
$10^5-10^7$ GeV (1 eV $<m_{3/2}<$ 10 keV).  (This scale depends on the
size of the messenger hypercharge coupling and the superpotential
couplings.)
The gravitino is the
lightest
supersymmetric partner, and the next to lightest supersymmetric
partner is a neutralino (linear combination of neutral gauginos,
higgsinos, and gauge singlet fermions) which should decay into a
photon and a gravitino with a lifetime in the range $10^{-13}-10^{-5}$
sec.
(Note that this decay rate is  more rapid than would be expected for
a process involving gravity because of the
goldstino component of the gravitino, and is so uncertain because it
is inversely
proportional to the fourth power of the supersymmetry breaking scale.)
This decay of the neutralino is a distinctive model independent signal
for
low energy supersymmetry breaking.
\item{6.} Some of the new particles predicted are potential dark matter
candidates. The R axion  is unstable once the R symmetry is broken;
for instance it
can decay into gravitino pairs.
The gravitino could provide an interesting amount of warm
dark matter if its mass is in the 10 keV range.  The massless charged fermion
of the supersymmetry breaking sector could be given a small mass
through higher dimension operators such as \eqn\dimfive{{\CO(1)\over m_P}
\bar U Q L E} and provide hot dark matter. The other particles of
the supersymmetry breaking sector can all decay into these. The $q$ and $\ell$
particles could decay by mixing with ordinary quarks and leptons, or
could be cold dark matter candidates. The other messenger particles
are all unstable since they   mix with the neutralinos and Higgs
scalars by a small amount.

\newsec{Conclusions}

We have presented a
renormalizable approach to spontaneous supersymmetry breaking, in
which all mass scales arise via dimensional transmutation.
Supersymmetric partner masses are calculable in terms of a few new
couplings, and at the weak scale the model resembles the MSSM, but
with a constrained parameter space. The absence of observed FCNC and
CP violation from the supersymmetry breaking sector is explained by
having flavor universal gauge couplings transmit supersymmetry
breaking to squarks, sleptons and gauginos. The supersymmetry breaking
sector can be as simple as an additional $SU(3)\times SU(2)\times U(1)$ gauge
theory with the
particle content of one family, and with communication of supersymmetry
breaking facilitated by a small
number of additional fields including vector like quarks and leptons and one
or
more gauge singlets. The only important role played by nonrenormalizable
supergravitional
couplings is to cancel the cosmological constant (by fine tuning) and
to give mass to an otherwise troublesome axion. The model is easily
made consistent with all terrestial, cosmological, and astrophysical
constraints. The lightest superpartner is the gravitino, which may
lead to a distinctive signal in future accelerators such as LEP II.
As one would expect for a dynamical model, these theories can
readily explain the hierarchy.  For example, in the 3-2 model,
with the assumption that all couplings are equal at the GUT
scale, the supersymmetry breaking scale is in the desired
$10^3$ TeV range.

Some readers may be concerned about our liberal
use of discrete symmetries, and the associated problem
of domain walls.  We do not view this problem as
 serious.  Our discussion does not require that
these symmetries be exact; if they are broken by
dimension five operators generated by Planck scale
physics, or by operators generated
at lower scales by gauge anomalies,
 these domain walls will quickly disappear.

In comparison with the conventional MSSM,  it is a great
advantage to have the supersymmetry breaking sector made explicit so
that supersymmetry breaking parameters are calculable.
The MSSM can arise from models in which supersymmetry is dynamically
broken in a gravitationally coupled ``hidden sector''. In fact, the
3-2 model which we have used as our prototypical example can be used
as a hidden sector model. Hidden sector models with dynamical
supersymmetry breaking have the virtue
 that they
can explain the origin of the hierarchy.  If, for example,
we take the two family $SU(5)$ model as hidden sector,
and assume that
the $SU(5)$ coupling is equal to the
unified coupling at $M_{GUT}$, we obtain
roughly $4 \times 10^{10}$ GeV for the SU(5) scale,
i.e. a quite reasonable intermediate scale value.
In addition, these models do not suffer from the conventional
Polonyi problem \ref\polonyi{G.D. Coughlan, W. Fischler,
K.W. Kolb, S. Raby and G.G. Ross, \pl{131}{1093}{59};
M. Dine, D. Nemeschansky and W. Fischler,
\pl{136}{1984}{169}.}
since the hidden sector, by assumption,
does not have flat directions\foot{Of course,
in the context of string theory, where one expects there
may be additional light moduli, there may
still be serious cosmological difficulties \ref\cosmo{T. Banks,
D.B.
Kaplan, A.E. Nelson,\physrev{49}{1994}{779}, B. de Carlos, A. Casas,
F. Quevedo and E. Roulet, \pl{318}{1993}{447}.}.}.
However there are still potential difficulties with the
hidden sector approach, such as the cosmological abundance of
gravitinos and flavor changing neutral currents,
which simply do not arise when supersymmetry breaking
occurs in a renormalizable visible sector theory. Thus
we feel that if nature turns out to be supersymmetric,
then the possibility of low
energy
supersymmetry
breaking should be taken seriously.

\centerline{\bf Acknowledgements}
The work of M. Dine was supported in part by the U.S. Department of
Energy.
The work of A. Nelson was supported in part by the DOE under grant
\#DE-FG06-91ER40614, by the NSF VPW program under grant
\#GER-9350061
 and by the Alfred P. Sloan Foundation.
A. N. would like to thank David Kaplan and Lisa Randall, the
organizers
of the ``Weak Interactions'' workshop at the ITP for their hospitality
during the inception of this project, and Shyamoli Chauduri for
asking whether a simpler dynamical supersymmetry breaking model
could be found.

\listfigs
\listrefs
\bye